\documentclass[twocolumn,showpacs,preprintnumbers,amsmath,amssymb]{revtex4}%
\usepackage{graphicx}
\usepackage{amsmath}
\usepackage{amsfonts}
\usepackage{amssymb}
\usepackage{hyperref}

\def\be{\begin{equation}}
\def\ee{\end{equation}}
\def\ba#1{\begin{array}{#1}}
\def\ea{\end{array}}
\def\bn{\begin{enumerate}}
\def\en{\end{enumerate}}

\def\summ{\sum\limits}

\providecommand{\U}[1]{\protect\rule{.1in}{.1in}}

\begin{document}
\preprint{APS/123-QED}
\title{Renormalization Group Approach to Oscillator Synchronization}
\author{Oleg Kogan$^1$, Jeffrey L. Rogers$^2$, M. C. Cross$^3$, G. Refael$^3$}
\affiliation{$^1$ Department of Materials Science, California
Institute of Technology, 1200 E. California Blvd., Pasadena, CA
91125}
\affiliation{$^2$ Control and Dynamical Systems, California
Institute of Technology, 1200 E. California Blvd., Pasadena, CA
91125}
\affiliation{$^3$ Department of Physics, California
Institute of Technology, 1200 E. California Blvd., Pasadena, CA
91125}

\date{\today}

\begin{abstract}
We develop a renormalization group method to investigate
synchronization clusters in a one-dimensional chain of
nearest-neighbor coupled phase oscillators.  The method is best
suited for chains with strong disorder in the intrinsic
frequencies and coupling strengths.  The results are compared with
numerical simulations of the chain dynamics and good agreement in
several characteristics is found.  We apply the renormalization
group and simulations to Lorentzian distributions of intrinsic
frequencies and couplings and investigate the statistics of the
resultant cluster sizes and frequencies, as well as the dependence
of the characteristic cluster length upon parameters of these
Lorentzian distributions.
\end{abstract}

 \maketitle

\section{Introduction}
\label{sec:Introduction} One of the most exciting aspects of
modern physics is the investigation of emergent collective
structures in many-body problems. Spontaneous synchronization in
networks of interacting nonlinear oscillators with different
(often randomly distributed) frequencies is one of the best
understood collective phenomena in nonequilibrium
systems~\cite{Synch book Pikovsky}. If the interaction between the
oscillators is weak, then the oscillators will evolve in an
uncoordinated fashion. In contrast, with proper connectivity and
coupling strength, the population can display the same frequency
as order forms. This collective behavior has been observed in
diverse systems including neural networks~\cite{Neural Networks1,
Neural Networks2, Neural Networks3}, Josephson
junctions~\cite{JJ1,JJ2,JJ3}, lasers~\cite{Lasers1}, electronic
circuits~\cite{Electronic Circuits}, and a range of biological
systems, including sleep cycles~\cite{Sleep_Cycles},
animal gaits \cite{Animal_Gaits} and
biological rhythms in general \cite{Biorhythms}.

Useful insights into the phenomenon of synchronization are
obtained by analyzing the idealized case of all-to-all coupling in
the limit of a large number of oscillators. In one of the most
common models each oscillator is described by a single variable,
phase, that advances at a constant rate in the absence of
coupling. This model is exactly soluble using a mean-field
approach, and it provides a framework for understanding
synchorinzation~\cite{Kuramoto,Kuramoto_review,Acerbon's_review}.
In this model, above a critical coupling strength the synchronized
phase appears: a finite fraction of oscillators evolve at a common
frequency. Many features of this phase, such as the growth of the
fraction of locked oscillators as a function of the coupling
strength, can be calculated.

In many physical systems finite range interactions are a more
realistic description. The simplest of these interactions are nearest
neighbor couplings~\cite{JJ4, Lasers2}. In general, finite range coupled
systems are much harder to analyze, and, as a result, there are few
theoretical tools that produce a quantitative description of the
collective motion. Much of our understanding of these systems rests on
numerical solutions of the governing equations. We propose
an alternative method to attack the synchronization problem. The method we
develop is a real-space renormalization group (RG) designed to work well
on systems with strong randomness, where both the intrinsic oscillation frequencies and coupling constants
are random variables taken from distributions with long tails. The
specific example we investigate contains Lorentzian distributions. This method is
shown to be accurate in systems with short range interactions.

In this paper, we analyze a one dimensional chain of oscillators
interacting with their nearest neighbors. Despite the relative simplicity
of the model, it can exhibit complex behavior since randomness
competes with a tendency to establish macroscopic order. This system is below
the lower critical dimension for a synchronization transition (Strogatz
and Mirollo~\cite{Strogatz and Mirollo 1, Strogatz and Mirollo 2} showed
that a nearest-neighbor chain cannot exhibit extensive synchronization),
and therefore can not be analyzed with mean-field techniques.
Collective behavior still arises, albeit at a finite length scale: the
chain becomes fragmented into finite length clusters of coherently moving
oscillators. We are interested in the distribution of cluster lengths,
cluster frequencies, and how these change as a function of the coupling
strengths and bare frequencies. Additionally, we seek to develop a
renormalization group scheme that will allow us to predict with some
accuracy the {\it individual} local behavior of oscillators. We also carry out an extensive numerical study of the problem and
use it as a benchmark for comparison with the RG.

The paper is organized as follows.  In Sec.
\ref{sec:Strong disorder RG} we outline the strong disorder
renormalization group approach with the details of the steps for the
nearest-neighbor oscillator chain derived in Sec. \ref{sec:RG
steps}. In Sec. \ref{sec:Results}, we compare some
key statistical features of the cluster state obtained by the RG with
results from direct numerical simulations of the equations of motion.
In Sec. \ref{sec:Physics} we outline our findings concerning the
physics of the synchronized clusters, presenting results for the
distributions of cluster sizes, cluster frequencies, as well as the
dependence of the characteristic length scale of the clusters on the
widths of the coupling and frequency distributions. Concluding
remarks are made in the last section.

\section{Strong Disorder Renormalization Group}
\label{sec:Strong disorder RG} To analyze the random oscillator
chain we develop a strong-disorder real space RG technique. A similar
method has been used successfully in quantum mechanical systems to
analyze the ground state of the random Heisenberg model
\cite{MaDas79,MaDas80,DSF94,DSF98}, and the superfluid-insulator
transition in random bosonic chains~\cite{AKPR04,AKPR08} (a
quantum reactive analog of the nearest-neighbor chain, where the
$\ddot{\theta}$ replaces $\dot{\theta}$, and $\theta$ is a quantum
operator).

The main idea behind the strong disorder approach applied to the
quantum-mechanical problem is best demonstrated in the case of the
spin-1/2 Heisenberg model:
\be
\hat{H}=\summ_i J_i \hat{\bf S}_i\cdot \hat{\bf S}_{i+1}.
\ee
The $i$'th term in the sum
considered on its own, splits the four states of the spins $i$ and
$i+1$, into a singlet state, and three triplet states excited by
the energy $J_i$. Thus a real-space decimation step can be
proposed. First, freeze the spin pair with the strongest interaction,
$J_n=\text{max}\{J_i\}$, into a singlet state. Perturbative
corrections to this state introduce an effective coupling between
the neighbors of the spins $n$ and $n+1$ which is:
\be
H_{RG}=\frac{J_{n-1}J_{n+1}}{2J_n} \hat{\bf S}_n\cdot \hat{\bf S}_{n+1},
\ee
an interaction identical in form to the operators in
the bare Hamiltonian, but with a much suppressed strength:
$J_{eff}=\frac{J_{n-1}J_{n+1}}{2J_n}<J_{n-1},\,J_{n},\,J_{n+1}$.
By repeating this decimation step, the number of free spins is
gradually decreased; at the stage when all spins are bound into
singlets, we obtain the ground state of the Hamiltonian.

The decimation step described above is justified if the coupling
to the rest of the chain is much weaker then between the two spins
in question. This is guaranteed if the chain considered has broad
coupling distributions. Below, we use a real-space renormalization
group approach to analyze the nearest-neighbor oscillator chain with
broad disorder distributions.

The nearest-neighbor oscillator chain presents a classical problem with
dissipative equations of motion, rather than a quantum problem with a
conserved Hamiltonian. Nevertheless, we can develop an analogous
strong disorder RG method for this problem as well. The chain is
governed by the equations of motion
\begin{equation}
\label{eq:model} \dot{\theta}_i = \omega_i +
K_{i-1}\sin{(\theta_{i-1} - \theta_i)} + K_i\sin{(\theta_{i+1} -
\theta_i)},
\end{equation}
where $\theta_i$ is the phase of the $i$th oscillator. The $\omega_i$
are the intrinsic frequencies taken from a random distribution (we
assume zero mean without loss of generality) and the $K_i$ give the
couplings to the nearest neighbors also taken from a random
distribution. The coupling will organize the oscillators into
clusters of common frequency $\bar{\omega}$ defined as
\begin{equation}
\label{eq:omega_bar_defn} \bar{\omega}_i \equiv \lim_{(t-t_0)
\rightarrow \infty} \frac{\theta_i(t) - \theta_i(t_0)}{t - t_0},
\end{equation}
such that each oscillator in a given cluster shares its
$\bar{\omega}$ with all the other oscillators in the same cluster. We
seek to understand the statistics of the cluster sizes and
frequencies.

The renormalization group approach is based on two observations. A
strong coupling between two oscillators tends to force them to rotate
at the same frequency forming a cluster that we can then treat as a
single effective oscillator. Concomitantly, oscillators
with anomalously large frequencies rotate essentially independently,
with little effect from their neighbors. These ideas lead to
two decimation steps: a strong coupling decimation and a
fast oscillator decimation. We apply these steps iteratively,
sweeping through the oscillators from large values of $\omega_i$ and
$K_i$ to smaller values.

\subsection{Decimation steps}
\label{sec:RG steps}
As in the case of the quantum Heisenberg chain, our real space RG will
be carried out iteratively by treating at each step the strongest term
in the right-hand-side of the equations of motion, i.e., the term with the highest
frequency. In the oscillator chain the strongest term at each step could be either a
strong coupling $K_i$, or a large frequency, $\omega_i$.
Let us now derive the appropriate decimation
steps for the nearest-neighbor chain for these two cases.  We begin by slightly
generalizing the model to
\begin{equation}
\label{eq:model_general} m_{i}\dot{\theta}_{i} = m_{i}\omega_{i} +
K_{i-1} \sin{(\theta_{i-1} - \theta_{i})} +
K_{i}\sin{(\theta_{i+1} - \theta_{i})},
\end{equation}
where a new parameter $m$ is added.  It will follow from the strong
coupling decimation step defined below that when two oscillators
with $m_{i}$ and $m_{i+1}$ are described by one effective
oscillator, its $m$ is given simply by $m_{i} + m_{i+1}$. Hence,
the value of $m$ represents the number of original oscillators
that this effective oscillator represents.  It will become
apparent that oscillators with larger $m$ are less affected by
perturbations from their neighbors, so $m$ is a measure of the
inertia of an oscillator and will be referred to as its ``mass'',
although it is not a true mass or inertia in the sense of Newton's
second law.

\subsubsection{Strong coupling decimation}
\label{sec:SC step}
\begin{figure}[h]
\begin{center}
\includegraphics[width=7cm]{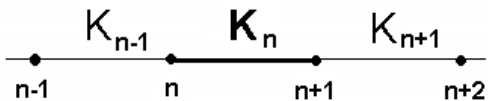}
\end{center}
\caption{\label{fig:strong K}A small piece of the chain around the
strongest coupling, labeled here as $K_n$.  All the neighboring
$K_i$ and all the local $\omega_i$ are assumed to have much
smaller magnitude then $K_n$.}
\end{figure}

The phases of two oscillators connected by a very large coupling
tend to advance in synchrony. The strong coupling step replaces
this pair of oscillators by a single effective oscillator with a
different mass and frequency determined as follows. Consider the
largest coupling in the chain: suppose this coupling is $K_{n}$,
Fig.~\ref{fig:strong K}. Because of the strong disorder the
frequencies $\omega_{n},\omega_{n+1}$ of the pair of oscillators
that $K_{n}$ couples and the couplings $K_{n-1},K_{n+1}$ to the
neighbors of the pair will almost always be small compared to
$K_{n}$. Ratios such as $\omega_{n}/K_n, K_{n-1}/K_n$, are therefore
expected to be small numbers; we will use a symbol
$\epsilon$ to denote quantities of this order. For small
$\epsilon$, $K_n$ will couple oscillators $n$ and $n+1$ into a
synchronized cluster that can be represented as a single effective
oscillator described by phase $\Theta$
\begin{equation}
\label{eq:strong_coupling_decimation}
\Theta\equiv(m_{n}\theta_{n}+m_{n+1}\theta_{n+1} )/M,
\end{equation}
possessing mass $M$,
\begin{equation}
\label{eq:strong_coupling_mass} M=m_{n}+m_{n+1},
\end{equation}
and an effective intrinsic frequency $\Omega$,
\begin{equation}
\label{eq:frequency_sum}
\Omega=(m_{n}\omega_{n}+m_{n+1}\omega_{n+1})/M.
\end{equation}
We refer to such pair of oscillators {\it strongly coupled}
(cf.~\footnote{Note that it is not the absolute magnitude of $K_n$ that
determines whether an oscillator is subjected to this step, but
the $O(\epsilon)$ ratios such as $\omega_{n}/K_n$ and
$K_{n-1}/K_n$: the phase difference between $\theta_n$ and
$\theta_{n+1}$  can not be assumed to be bounded, even if the
coupling $K_n$ between them is very large, when the difference
between their intrinsic frequencies or another neighboring $K_n$
are comparably large.} and Sec.~\ref{sec:RG_Implement}).

The small frequency difference $\omega_{n+1}-\omega_{n}$ and the
small couplings $K_{n-1}$ and $K_{n+1}$ induce a small phase
difference $\delta\equiv\theta_{n+1}-\theta_{n}$ between the two
oscillators. Using $\theta_{n}=\Theta-m_{n+1}\delta/M$ and
$\theta_{n+1}=\Theta+m_{n}\delta/M$, Eqs.~(\ref{eq:model_general})
gives the evolution equation for the phase of
the new effective oscillator%
\begin{eqnarray}
\label{eq:strong_coupling_Theta}
M\dot{\Theta} = M\Omega &+& K_{n-1}\sin{(\theta_{n-1}-\Theta+\frac{m_{n+1}}{M}%
\delta)} \nonumber \\
 &+& K_{n+1}\sin(\theta_{n+2}-\Theta-\frac{m_{n}}{M}\delta
).
\end{eqnarray}
There are corresponding equations for $\dot{\theta}_{n-1}$ and
$\dot{\theta}_{n+2}$.  Aside from the new $\delta$-terms, the
equations take the original form.

There are two contributions to the phase $\delta$.  The
contribution from the frequency difference is a time independent
phase given to leading order in $\epsilon$ by
\begin{equation}
\delta_{0}=\frac{\mu_{n,n+1}}{K_{n}}(\omega_{n+1}-\omega_{n}).
\end{equation}
where $\mu_{i,j}^{-1}=m_{i}^{-1}+m_{j}^{-1}$ is the reduced mass. This
can be seen by considering just the two oscillators coupled by the strong
$K_n$ and disregarding effects of the rest of the chain.  Then, as can
be seen from the difference of their equations of motion,
Eqs.~(\ref{eq:model_general}), as long as $\omega_{n+1} - \omega_n$ is
smaller then $K_n/\mu_{n,n+1}$, which will be the case if $K_n$ is much
larger then all the neighboring parameters, these two oscillators
will reach a state in which they rotate at the same frequency and are
separated by the fixed phase difference $\delta_0$.  We will see this
explicitly after writing down an equation of motion for $\delta$
below. For the one dimensional chain we can redefine $\theta_{n-1}$
and prior phases, as well as $\theta_{n+2}$ and subsequent phases, to
absorb such constant phases, and so we may ignore them.  This
maneuver is restricted to one dimension in the spatial lattice, and
the extension of the method to higher dimensions would require a more
careful study of the effect of these additional phases.

The couplings $K_{n-1}$ and $K_{n+1}$ to the neighbors of the pair
give an additional contribution to $\delta$ that is time dependent
as the neighboring oscillators evolve. To leading order in $\epsilon$
it turns out that these oscillating corrections give small extra
coupling terms of forms not included in the equation of motion
Eqs.~(\ref{eq:model_general}): a second nearest neighbor interaction
between oscillators, a three body interaction involving the new
effective oscillator and the neighboring pair, and a second harmonic
correction to the form of the pairwise interaction.  The detailed forms
of these terms can be found in~\cite{Oleg_Thesis}. We will neglect these
more complicated interaction terms since they are much weaker then all
other competing interactions.

These results are derived from the exact expression for the
equation of motion of the phase difference $\delta(t)$,
\begin{align}
\dot{\delta} &
=-\frac{K_{n}}{\mu_{n,n+1}}\sin{\delta+(}\omega_{n+1}-\omega
_{n})\nonumber\\
&  +\frac{K_{n+1}}{m_{n+1}}\sin{\left(  \theta_{n+2}-\Theta-\frac{m_{n}}%
{M}\delta\right)  }\nonumber\\
&  +\frac{K_{n-1}}{m_{n}}\sin{\left(  \theta_{n-1}-\Theta+\frac{m_{n+1}}%
{M}\delta\right)}.
\end{align}
This equation describes a damped particle in a time dependent washboard
potential. The relaxation rate $K_{n}/\mu_{n,n+1}$ given by the first term on
the right hand side is fast compared with the other terms in the
equation. The solution can therefore be developed as an expansion in
a ratio of time scales, and the leading order term is the adiabatic
solution
\begin{multline}
\delta\simeq\frac{\mu_{n,n+1}}{K_{n}}\left[  (\omega_{n+1}-\omega_{n})+\frac{K_{n+1}%
}{m_{n+1}}\sin{\left(  \theta_{n+2}-\Theta\right)  }\right. \\
+\left.  \frac{K_{n-1}}{m_{n}}\sin{\left(
\theta_{n-1}-\Theta\right) }\right].
\end{multline}
The first term in the braces gives the constant phase
$\delta_{0}$, and on substituting back into
Eq.~(\ref{eq:strong_coupling_Theta}) and expanding to first order
in small $\delta$, the second and third terms give the corrections
to the interaction terms that we will neglect, as discussed above.
Specifics of the calculations outlined in this section are
presented in~\cite{Oleg_Thesis}.

In summary, the effect of the strong coupling decimation step is
to replace the pair of oscillators with phases $\theta_{n}$ and
$\theta_{n+1}$ by a single oscillator with phase $\Theta$, mass
$M$ and frequency $\Omega$ given by
Eqs.~(\ref{eq:strong_coupling_decimation}),
(\ref{eq:strong_coupling_mass}), and~(\ref{eq:frequency_sum}).

\subsubsection{Fast oscillator decimation}
\label{sec:CO step}

\begin{figure}[tb]
\begin{center}
\includegraphics[width=8cm]{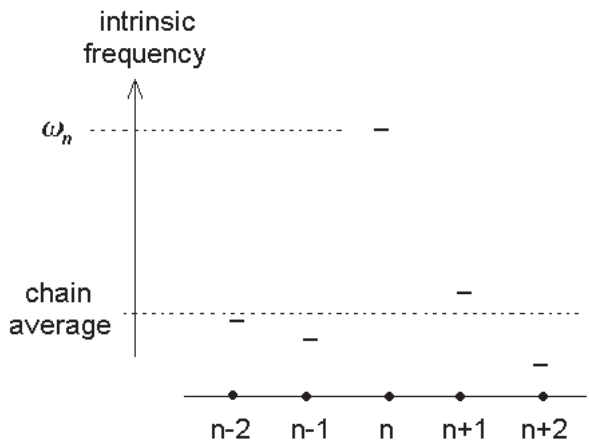}
\end{center}
\caption{\label{fig:CO}A small piece of the chain around the
oscillator with the largest frequency, labeled here as
$\omega_{n}$. All the neighboring $K_i$ and all the neighboring
$\omega_i$ are assumed to have much smaller magnitude then
$\omega_n$.}
\end{figure}

Next we consider the case where the highest frequency scale in the
problem is a bare frequency. In this case we apply the notion that
the phase of an oscillator with frequency much different from its
neighbors tends to advance as if the oscillator were decoupled.
Consider the oscillator with the largest frequency in the entire
chain: suppose this is oscillator $n$ with intrinsic frequency
$\omega_n$, Fig.~\ref{fig:CO}.  Because of strong disorder the
coupling strengths $K_{n-1}$ and $K_{n}$ to the neighboring
oscillators and the frequencies $\omega_{n\pm 1}$ of these
oscillators will almost always be small compared to $\omega_n$.
Therefore we expect ratios such as  $K_{n-1}/\omega_n$ and
$\omega_{n-1}/\omega_n$ to be small compared to 1; as before we
denote quantities of this order by $\epsilon$.  To zeroth order in
$\epsilon$ the $n$'th oscillator runs freely, with the frequency
$\omega_n$. We call such oscillators ``fast oscillators'' (see
\footnote{Similarly, it is not the absolute magnitude of
$\omega_n$ that determines whether an oscillator is subjected to
this step, but the $O(\epsilon)$ ratios such as $K_{n-1}/\omega_n$
and $\omega_{n-1}/\omega_n$: the phase of an oscillator with a
large $\omega_n$, but strongly coupled to a slow neighbor or
positioned next to another fast neighbor can not be assumed to
advance freely.} and Sec.~\ref{sec:RG_Implement} for elaboration).

Deviations from the free-running solution can be calculated
perturbatively in $\epsilon$.  The most important effect we might
look for is an induced effective coupling via oscillator $n$
between the neighbors $n-1$ and $n+1$, since this might lead to
their synchronization. However, at least to order $\varepsilon^2$
as we describe below, we find that there is no induced interaction
that tends to lock the neighboring oscillators \footnote{This
should be compared with the strongly disordered spin chain, where
an effective interaction between the neighbors is given by
eliminating the high energy spins.}.  Thus the main effect of
decimating the largest frequency oscillator in the chain is to
eliminate this oscillator from further consideration (it
represents a cluster in the final tally at frequency close to
$\omega_n$ and with a size given by its mass parameter $m$) and to
cut the chain into two pieces. This is a key feature of the
strongly random chain that limits the size of synchronized
clusters.  At order $\epsilon^2$ we do find a renormalization of
the frequencies of both the eliminated oscillator and the two
neighbors remaining in the chain.

We now investigate the perturbation of the motion of the fast
oscillator (oscillator $n$ in Fig.~\ref{fig:CO}) given by
Eqs.~(\ref{eq:model_general}).  To zeroth order in $\epsilon$, the
solution to $\theta_n$ is given by the uniform running state
$\theta_n=\omega_n t$.  There are two $O(\epsilon^2)$ effects that
slightly perturb the average rate of this uniform advancement of
the phase.  The first is the fact that oscillator $n$ rotates in
the tilted washboard potential created by its neighbors.  Since
typically $\omega_{n-1},\omega _{n+1} \ll \omega_n$ for strong
disorder, this potential can be treated as effectively static. The
second effect is a small component of fast dynamics added to the
motion of the neighboring oscillators, which then acts back on
oscillator $n$. Both of these effects change the average frequency
of the eliminated oscillator by a shift of order $\epsilon^2$.
There are corresponding corrections to the motion of the
neighboring oscillators as well. These effects will be apparent in
the following analysis.


First, consider the motion of $\theta_{n}$ in the
nearly static washboard potential from the neighbors, calculated
to second order in the coupling. Setting the neighboring phases in
Eqs.~(\ref{eq:model_general}) to fixed values
$\theta_{n\pm1}^{(0)}$ gives the oscillation period of the $n$th
oscillator.
\begin{equation}
\label{eq:Period}
T_{n}=\int_{0}^{2\pi}\frac{d\theta_{n}}{\omega_n + \frac{K_{n-1}}{m_{n}}%
\sin{(\theta_{n-1}^{(0)}-\theta_{n})}+\frac{K_{n}}{m_{n}}\sin{(\theta
_{n+1}^{(0)}-\theta_{n})}}.
\end{equation}
Expanding the denominator and performing the integration over
$\theta_{n}$ gives the change to the frequency of the $n$th
oscillator $\bar{\omega}_n - \omega_n$ with $\bar{\omega}_n = 2\pi
T_{n} ^{-1}$
\begin{equation}
\label{eq:Delta_1}
\delta\bar{\omega}_n^{(1)}\simeq-\frac{K_{n-1}^{2}}{2m_{n}^{2}\omega_n}-\frac{K_{n}^{2}%
}{2m_{n}^{2}\omega_n}-\frac{K_{n-1}K_{n}}{m_{n}^{2}\omega_n}\cos{(\theta
_{n+1}^{(0)}-\theta_{n-1}^{(0)})}.
\end{equation}

The second effect perturbing the motion of the fast oscillator arises
from the rapid induced motion of the neighboring oscillators
$\theta_{n\pm1}\simeq\theta_{n\pm1}^{0}+\delta\theta_{n\pm1}$. For
example from Eqs.~(\ref{eq:model_general}) for oscillator $n-1$ we have%
\begin{align}
\label{eq:delta_theta_n-1} \delta\theta_{n-1} &
\simeq\frac{K_{n-1}}{m_{n-1}}\int dt~\sin{(\omega_n
t-\theta_{n-1}^{0})}\\
&  {=}{-}\frac{K_{n-1}}{m_{n-1}\Omega}\cos{(\omega_n
t-\theta_{n-1}^{0}),}
\end{align}
where we use the zeroth order solution for oscillator $n$ and the
influence of the next nearest neighbor $\theta_{n-2}$ only appears
at higher order. This provides a perturbation back on the $n$th oscillator
given by expanding the second term on the right hand side of
Eqs.~\ref{eq:model_general} for $i=n$
to first order in $\delta \theta_{n-1}$%
\begin{equation}
\delta\dot{\theta}_{n} \simeq \frac{K_{n-1}}{m_{n}}\cos{(\omega_n
t-\theta_{n-1}^{0})}\delta \theta_{n-1}.
\end{equation}
Using Eq.~(\ref{eq:delta_theta_n-1}) for $\delta\theta_{n-1}$,
averaging over the fast oscillations, and adding the similar
effect from oscillator $n+1$ gives
the change in the average frequency of the $n$th oscillator%
\begin{equation}
\label{eq:Delta_2}
\delta\bar{\omega}_n^{(2)}\simeq-\frac{K_{n-1}^{2}}{2m_{n}m_{n-1}\omega_n}-\frac
{K_{n}^{2}}{2m_{n}m_{n+1}\omega_n}.
\end{equation}
Combining Eqs.~(\ref{eq:Delta_1}) and (\ref{eq:Delta_2}) gives the
total renormalization
of the frequency of the fast oscillator to second order%
\begin{multline}
\delta\bar{\omega}_n \simeq
-\frac{K_{n-1}^{2}}{2m_{n}\mu_{n,n-1}\omega_n}-\frac{K_{n}
^{2}}{2m_{n}\mu_{n,n+1}\omega_n} \\ 
-\frac{K_{n-1}K_{n}}{m_{n}^{2}\omega_n}\cos{(\theta_{n+1}^{{}}-\theta_{n-1})},
\label{eq:combined}
\end{multline}
with $\mu_{i,j}^{-1}=m_{i}^{-1}+m_{j}^{-1}$ as before.

There are reciprocal effects on the neighboring oscillators that can
be calculated by a similar procedure. Since at the pairwise level
interactions between the oscillators conserve the mass-weighted
frequency average \footnote{By adding all equations of motion we can
see that $\sum_{i=1}^N m_i \delta \omega_i = 0$; this is just a
consequence of the fact that interactions are odd. Moreover, at the
level of approximation that we consider, $\delta \omega_i = 0$ for
all but the fast oscillator $n$ and its nearest neighbors, $n-1$ and
$n+2$.  Hence $m_{n-1} \delta \omega_{n-1} + m_n \delta \omega_n +
m_{n+1} \delta \omega_{n+1} = 0$.}, we can immediately see that the
first two terms in Eq.~(\ref{eq:combined}) correspond to frequency
renormalizations of the neighboring oscillators
\begin{equation}
\label{eq:neighbor_corrections}
\delta\bar{\omega}_{n-1}=\frac{K_{n-1}^{2}}{2m_{n-1}\mu_{n,n-1}\omega_n}
,\ \delta
\bar{\omega}_{n+1}=\frac{K_{n}^{2}}{2m_{n+1}\mu_{n,n+1}\omega_n}.
\end{equation}

The fact that the first two frequency correction terms in
Eq.~(\ref{eq:combined}) are additive suggests that they can be
obtained from the analysis of a pair of 2-oscillator systems. This
calculation can be performed exactly for the phase difference, for
example $\phi = \theta_n - \theta_{n-1}$.  Such a frame independent
analysis reproduces at $O(\epsilon^2)$ the first two terms in
Eq.~(\ref{eq:combined}) and the terms in
Eq.~(\ref{eq:neighbor_corrections}) which were obtained in the
specific reference frame of the average frequency, but also
yields terms that are higher order in $\epsilon$. In the context
of the full chain including these higher order terms is an uncontrolled
approximation, however empirically we find them to produce a better match with
the simulation data in some rare cases.  Hence, in the
numerical renormalization of the chain, we employ the expressions
produced by this analysis.

The phase difference of oscillators $n$ and $n-1$ isolated from
the rest of the chain satisfies
\begin{equation}
\dot{\phi} = (\omega_n - \omega_{n-1}) -
\frac{K_{n-1}}{\mu_{n,n-1}}\sin{\phi}.
\end{equation}
A period $T$ over which this phase difference grows by $2\pi$ can
be defined analogously to Eq.~(\ref{eq:Period}), but since there
is only one sine term, the integral can be computed analytically.
From that solution and the fact that for the isolated pair $m_1 \dot{\theta}_1 + m_2
\dot{\theta}_2 = m_1 \omega_1 + m_2 \omega_n$ we obtain the
correction to the frequency of oscillator $n$ coming from oscillator $n-1$:
\begin{multline}
\label{eq:sq_root_formula} \delta \bar{\omega}_{n,n-1} = \\
\frac{\mu_{n,n-1}(\omega_{n-1} - \omega_n)}{m_n}\left[1 -
\sqrt{1-\left(\frac{K_{n-1}}{\mu_{n,n-1}(\omega_n -
\omega_{n-1})}\right)^2}\right].
\end{multline}
We use this expression to replace the first term on the right hand side of
Eq.\ (\ref{eq:combined}) with a corresponding expression for the first
term in Eq.\ (\ref{eq:neighbor_corrections}),
and similar expressions for the second pair of terms.

The new cosine interaction term in Eq.\ (\ref{eq:combined}) also
warrants discussion.  There is a corresponding renormalization of
the equations of motion of each of its neighbors.  In the motion
of the fast oscillator Eq.~(\ref{eq:combined}), we ignore the
cosine term since it averages to zero on the long time scale of
the period of oscillators $n\pm1$.  In the equations for
$\dot{\theta}_{n-1}$and $\dot{\theta}_{n+1}$  the corresponding
term is a coupling $\propto \cos{(\theta_{n+1}-\theta_{n-1})}$
between these oscillators, with the same sign in the equations of
motion of both neighboring oscillators.  Therefore this
interaction does not have action-reaction symmetry. This sign
peculiarity arises because the decimated oscillator has a
particular direction of rotation. Because of the lack of
action-reaction symmetry, the cosine term does not tend to pull
neighboring oscillators together, and so does not directly
influence the clustering. We therefore omit this term in the
renormalization procedure.

\subsection{Implementation}
\label{sec:RG_Implement} The chain of oscillators is renormalized
by successive application of the two decimation steps developed
above.  These are executed numerically on a list of parameters
$(m_i, \omega_i, K_i)$ representing the chain of oscillators. A
single decimation step consists of finding the largest frequency
of the set $\left\{\omega_i, \frac{K_i}{2\mu_{i,i+1}}\right\}$,
renormalizing the appropriate term as described above, and working
down to smaller ones later.  Notice that it is the largest
$\frac{K_i}{2\mu_{i,i+1}}$ that identifies the pair of oscillators
to be subjected to the strong coupling step, not the largest
$K_i$. The reason for this becomes clear when all equations are
divided by their respective $m_i$.  An equation for the phase
difference $\theta_{i+1} - \theta_i$ contains the term
$\frac{K_i}{\mu_{i,i+1}}\sin{(\theta_{i+1} - \theta_i)}$. When
$\frac{K_i}{\mu_{i,i+1}}$ is much larger then the terms
representing coupling to the neighbors, this phase difference will
tend to be locked and the strong coupling step should be applied
to the pair of oscillators $i$ and $i+1$. The factor of $2$ was
used in the denominator to make sure that the initial stages of
the RG procedure, couplings and frequencies are treated on equal
footing.

Initial values of masses are all set to $1$.
The numerical procedure identifies the oscillators with $\omega$ and $K$ that
lie in a narrow band of magnitudes at the top of the spectrum of
remaining values, and then decimates those oscillators according to
the steps defined in Sec.s \ref{sec:SC step} and \ref{sec:CO
step}.  A band is used, rather than just selecting the largest
values, to improve the efficiency of the code. The width of the band
is chosen to be narrow ($1\%$ of the chain) to maintain the descending
order in $\omega$ and $K$ for the decimation of nearby oscillators.
Oscillators decimated in the fast oscillator step form a cluster in the final
tally, with the effective mass describing the size of the cluster
and the frequency giving the
time-averaged rate at which the phase of this cluster advances. As
described in Sec. \ref{sec:CO step}, these oscillators are
removed from the chain.  After the decimation of oscillator band, the remainder of the chain consists of fewer
oscillators, which have lower coupling strengths and frequencies
within a narrowed spectrum. The procedure is then repeated for successively
lower bands of $K$ and $\omega$.

During the RG execution, two types of special cases related to
violations of the assumptions based on strong disorder are occasionally
encountered.  First, in our treatment of the
oscillator with the largest $\omega$ we have made an assumption
that nearby $\omega$ and $K$ are small in comparison.
This assumption defines the
physical content of the strong disorder assumption.  In real chains,
it is possible to have neighbors with parameters which are
somewhat lower 
yet the difference is not large enough to guarantee that effects
of these neighbors 
are small.  This is especially important as the spectrum of the
remaining $K$ and $\omega$ values shrinks.  To be more precise,
for a decimation of a fast oscillator $n$, the magnitude of $r_{FO}$
defined by
\begin{equation}
r_{FO} \equiv \mathrm{Max}\left[ \frac{K_n}{\mu_{n,n+1}|\omega_n -
\omega_{n+1}|},
\frac{K_{n-1}}{\mu_{n,n-1}|\omega_n - \omega_{n-1}|}\right]
\end{equation}
determines whether the effects of the neighbors are
small, so that the step can be performed.
Similarly, for a strong coupling decimation of oscillators $n$ and
$n+1$,  the magnitude of the ratio
\begin{equation}
r_{SC} \equiv \frac{K_n}{\mu_{n,n+1}|\omega_{n} - \omega_{n+1}|}
\end{equation}
determines whether the step should be performed. In our
implementation of the RG we set set the criteria to $r_{FO}<1$ and
$r_{SC}>1$. This choice was based on the numerical study of small
chains ($N < 10$) as well as on the fact that for sinusoidal
coupling $r_{FO}$ does not need to be very small for $\omega_n t$
to give a good approximation to the dynamics of the $n$th
oscillator. This is hinted by the square-root dependence of the
frequency correction in Eq.~(\ref{eq:sq_root_formula}): analysis
of that formula indicates that $\bar{\omega}_n$ quickly approaches
$\omega_n$ as $r_{FO}$ decreases below $1$.  A more thorough
discussion of the choice for $r_{FO}$ and $r_{SC}$ can be found in
\cite{Oleg_Thesis}.

There is a second type of violation of the assumptions based on
the strong disorder.  Note that single oscillators with $m=1$ or
oscillators representing clusters with $m>1$ are normally
subjected to the fast oscillator decimation step at some point in
the RG procedure: most of the bare oscillators either join other
oscillators in building clusters, which are eventually decimated
via the fast oscillator step, or they are decimated as fast
oscillators directly without undergoing a strong coupling step.
But some rare combinations of frequencies and
couplings will prevent the decimation from occurring at all within
this scheme.
For example, while an oscillator's frequency may lie in the
executable band, this oscillator may be coupled via a $K$ which is
outside this band (it has a somewhat lower magnitude), but may
nevertheless cause $r_{FO}$ to exceed $1$.  In general, when the RG reaches the
scale of that coupling, it will perform a strong coupling
decimation step.  However, in rare cases this may not happen due
to intervening steps that have
modified the neighboring parameters, rendering that coupling no
longer strong according to the $r_{SC}$ criterion.  As a result,
the site in question has not been subjected to either decimation
step. To address such unusual cases, the algorithm sweeps repeatedly
through the entire spectrum of $\omega$ and $K$ values until nothing is left to
decimate.

\section{Numerics}
\label{sec:Results}

While the discussion above constructed a general RG algorithm, let
us now explore the scheme within a disorder realization with a
specific class of frequency and coupling distributions
$g(\omega),\,G(K)$. The bare frequencies in the system we analyze
have a symmetric Lorentzian distribution with a cutoff:
\begin{equation}
\label{eq:g(w)} g(\omega) = \frac{C_1}{1 + \omega^2},
\end{equation}
with $-\omega_{c}\leq \omega \leq \omega_{c}$. The couplings are
positive and are taken from a half Lorentzian with a cutoff:
\begin{equation}
\label{eq:G(K)} G(K) = \frac{2\mu C_2}{\mu^2 + K^2}
\end{equation}
with $0\leq K \leq K_{c}$.  The cutoffs are used to facilitate the
effective numerical simulations (Sec.~\ref{sec:num_sim} below):
increasing cut-off values increases the time needed for
simulations.   The same cut-off values for $\{\omega\}$ and
$\{K\}$ were chosen: $\omega_c = K_c = 100$; 
the values of these cutoffs are much larger then the FWHM of all
Lorentzians that we considered. The $C_1$ and $C_2$ are
normalization constants that become $1/\pi$ when the cut-off
values $\omega_c$ and $K_c$ are infinite. Notice that due to the
structure of Eqs.~(\ref{eq:model_general}), it is always possible
to divide all equations by a constant such that the width of one
of these distributions will be unity: this only changes the time
scale of all the processes in the chain.
We chose $g(\omega)$ to have unity width and explore the physics
by varying the other width parameter $\mu$.

\subsection{Renormalization group}
\label{sec:RG_details} The RG procedure is applied to a chain of
$10^6$ oscillators.  Twelve values of coupling width $\mu$ are
studied: $\mu=0.025$, $0.05$, $0.25$, $0.5$, $0.625$, $1.25$,
$2.5$, $3.75$, $4.5$, $5$, $6.25$, and $7.5$.  The set of random
numbers defining the particular realizations of the chain were
different in the RG and in the simulations (below) for the
statistical comparisons, but were identical for the real-space
comparison described in Sec.~\ref{sec:Analysis}.

\subsection{Simulations}
\label{sec:num_sim} In order to explore both the system's
characteristics and the reliability of the renormalization group,
Eqs.~(\ref{eq:model}) are also integrated numerically. The
numerical method is a variable stepping Runge-Kutta algorithm.
Systems of $N=10000$ oscillators are solved with $N$ intrinsic
frequencies chosen randomly from the cutoff distribution given by
Eq.~(\ref{eq:g(w)}). Similarly, $N-1$ coupling constants are
randomly selected from the cutoff distribution given by
Eq.~(\ref{eq:G(K)}).  The same twelve values of coupling width
$\mu$ that are used in the RG are studied in the simulations.
Results for each value of $\mu$ are averaged over $100$
realizations of the intrinsic frequencies and coupling constants.
To help facilitate comparisons between different distribution
widths the same 100 random number
seeds are used for each distinct coupling width.  

The simulations are performed for a relatively long time of
$t=10000$.  Running phases at each site are recorded at regular
intervals and used to calculate average frequencies over a time
$T$ according to
\begin{equation}
\label{eq:omega_bar_numerics} \bar{\omega}_{i}(T) =
\frac{\theta_{i}(t_{0}+T) - \theta_{i}(t_0)}{T};
\end{equation}
with  $T=t-t_0$ [compare with the theoretical definition of $\bar{\omega}$
given by Eq.~(\ref{eq:omega_bar_defn})]. The time $t_0$
is chosen to eliminate transients:  increasing values
are tried until average frequencies remain unchanged.  In
these simulations, the value of $2\pi/T$ sets the
resolution limit in $\bar{\omega}_{i}$ to distinguish two neighboring
frequency clusters.  A group of oscillators is determined to be
members of a synchronized cluster 
only if all the members ($i=n, n+1, \ldots, n+m+1$) have the same
value of $\bar{\omega}_{i}$ as the neighboring sites within some
tolerance $\left|\bar{\omega}_{\ell} -
\bar{\omega}_{i}\right|\leq\eta$ $(\ell=i-1$ and $i+1)$. Since
$T^{-1}$ is $O\left(10^{-4}\right)$ the
tolerance $\eta$ is set to $10^{-3}$.

\subsection{Comparison of RG and simulations.}
\label{sec:Analysis}
\begin{figure*}
\begin{center}
\includegraphics[width=17.7cm]{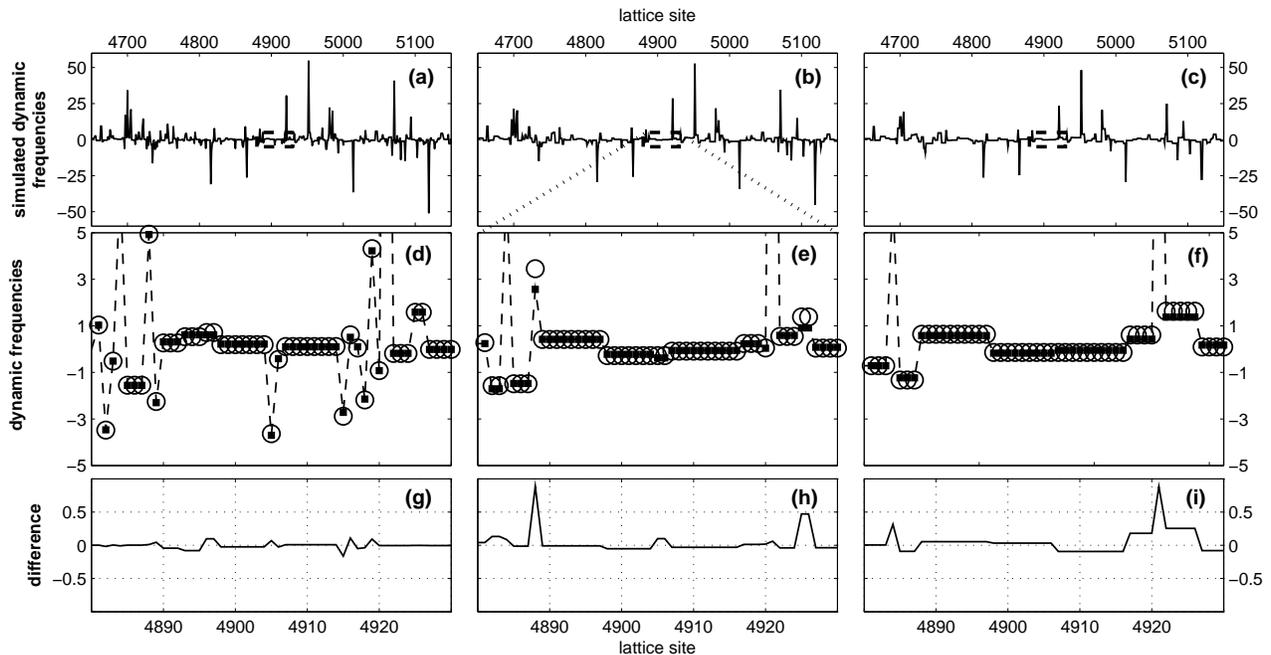}
\end{center}
\caption{\label{fig:real_space_comparison} Comparisons between
cluster structure of dynamic frequencies found in numerical
simulations of a 10000-oscillator chain with those predicted by RG
over a representative number of lattice sites. The top row (a-c)
plots the dynamic frequencies found for 500 oscillators in the
simulations and demonstrate how some oscillators maintain large
frequencies. The dashed boxes in the centers of (a-c) encircle 50
oscillators magnified in panels (d-f), respectively. In (d-f)
average frequencies from simulations given by
Eq.~(\ref{eq:omega_bar_numerics}) are denoted by solid squares,
while RG predictions are shown as open circles. The differences
between RG and simulations are shown in panels (g-i). Each column
shows results for a different coupling width: $\mu=1.25$ (a,d,g),
$\mu=3.75$ (b,e,h), and $\mu=7.5$ (c,f,i).}
\end{figure*}

While we are primarily interested in correctly capturing the
statistical characteristics through RG, the local validity of our procedure
can be established by comparing the frequency cluster structure that it
predicts with results from the simulations.  
Figure~\ref{fig:real_space_comparison} plots these comparisons
over a representative sample of 150 oscillators from chains of
$N=10000$ at 3 different coupling distribution widths.
For all 3 values of $\mu$, there are clusters with large frequency
oscillators interspersed. As expected, the characteristic size of
these clusters grows with $\mu$. 
There is excellent agreement of the cluster sizes, locations,
and frequencies between the two approaches. This provides some
confidence that the essential mechanisms are captured by the RG
procedure.

Analysis of these and similar plots, shows that the agreement
of the cluster frequencies becomes more accurate
with increasing cluster size. The reason for this result is the
smaller boundary-to-bulk ratio of larger clusters, which is more
apparent if Eqs.~(\ref{eq:model_general}) is rewritten as
\begin{equation}
\label{eq:model_alternative} \dot{\theta}_i = \omega_i +
\frac{K_{i-1}}{m_i}\sin{(\theta_{i-1} - \theta_i)} +
\frac{K_i}{m_i}\sin{(\theta_{i+1} - \theta_i)}
\end{equation}
As clusters become larger, the importance of the coupling to the
rest of the chain is weighted down by a $1/m$ factor.  In the fast
oscillator decimation, this is manifested by the appearance of
this factor in the frequency correction. Also, when a cluster is
built up by a strong coupling decimation step, any frequency
renormalizations of the component oscillators by prior fast
oscillation decimation steps cancel when the weighted sum of the
frequencies is calculated to give the frequency of the new
cluster. Thus the cluster frequencies $\bar{\omega}_i$ are given
to a good approximation by the average of the constituent bare
$\omega$ values, plus the small correction due to the fast
oscillator decimation step that yields the final cluster.

Investigating the frequency distribution of clusters of a given
size provides further insight into the validity of this RG
procedure.
\begin{figure*}
\begin{center}
\includegraphics[width=16.5cm]{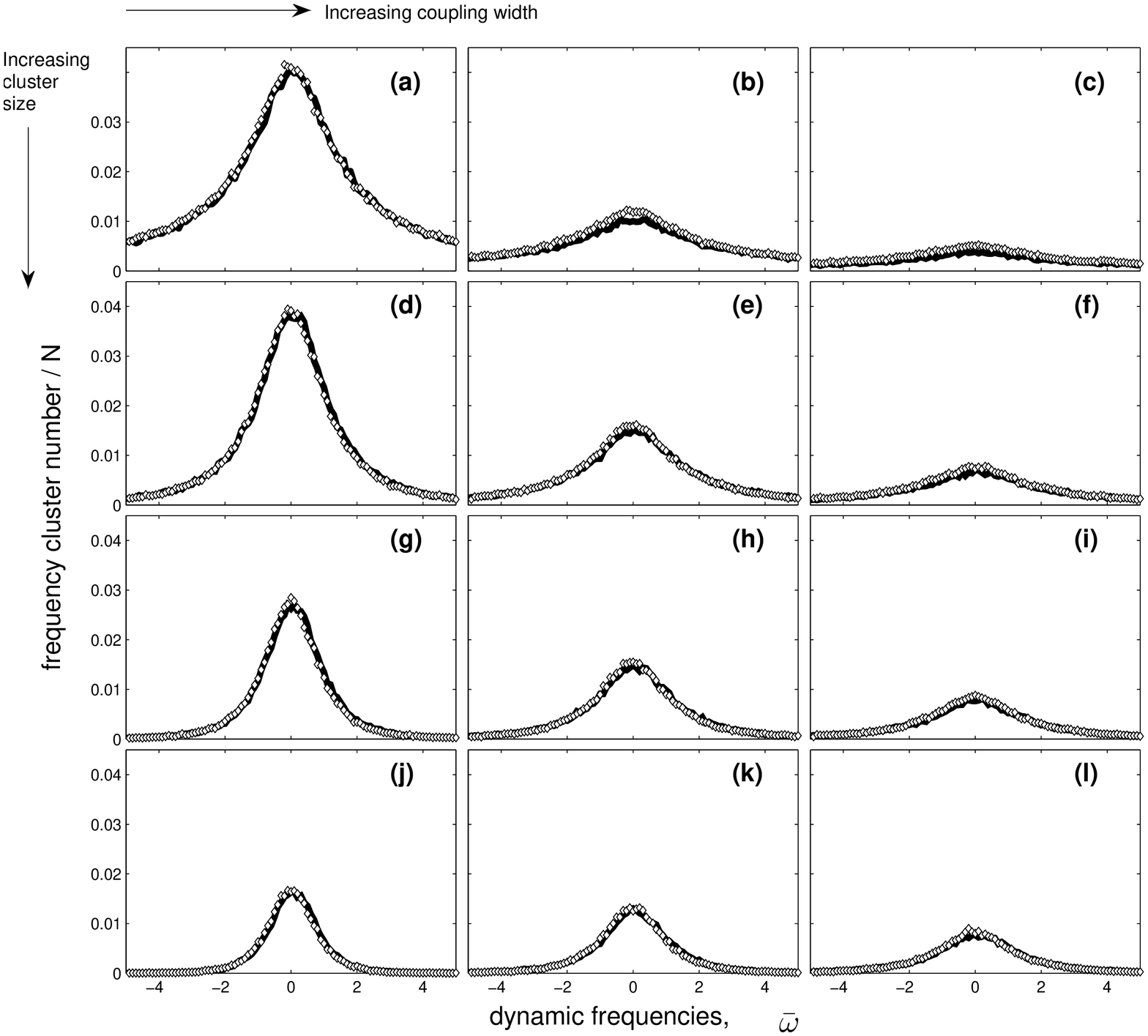}
\end{center}
\caption{\label{fig:Freq_hist} Number of clusters of a particular
  frequency divided by the total length of the chain ($N=10^6$). The
  plot compares the RG prediction (solid lines)
and the simulation results (open diamonds) for 3 coupling widths and
4 cluster sizes. Each panel has a fixed coupling width and
cluster size. Frequencies, on the hortizontal axes, are grouped
into bins of size $\Delta\bar{\omega} = 0.1$. The rows are each a
different cluster size; $m = 1$ (a-c), $2$ (d-f), $3$ (g-i), and
$4$ (j-l) from top to bottom. Each column is a different coupling
width: $\mu=1.25$ left, $\mu=3.75$ center, and $\mu=7.5$ right.
}
\end{figure*}
This is shown in Fig.~\ref{fig:Freq_hist}, where for each cluster
size $m$ from 1 to 4, the RG predictions are compared with the
results from simulations using three values of the parameter
$\mu$. The two methods are in excellent agreement for the larger values of $m$
and also for all $m$ at small values of $\mu$.

The comparisons in Fig.\ \ref{fig:Freq_hist} show a small but
systematic underestimation of the number of smallest sized
clusters ($m=1$
and $m=2$) near zero frequency and large coupling [panels (b) and (c) and less so in in panels (e) and (f)]. 
The rms difference between the RG generated and numerically
calculated curves ranges between $0.7 - 2.2\cdot 10^{-4}$, with the
worst agreement in panel (b), and best agreement in panel (l).
The region of the greatest discrepancy, small $m$ and $\omega$, is
in fact where the fast oscillator decimation is expected to be
least accurate, since by the time the RG sweep reaches these
frequencies, the distributions of remaining frequencies and
couplings are no longer expected to be wide so that the
approximations based on strong randomness are not as accurate.
However the disagreement decreases with increasing $m$, because,
as described above, the frequencies of larger clusters are
insensitive to the corrections from neighbors.

\section{The unsynchronized phase of an oscillator chain}
\label{sec:Physics} 

Let us now use the results of the RG and the simulations to discuss the
statistical properties of the frequency clustering in the one dimensional
oscillator chain with strong randomness.

\subsection{Cluster size distribution}
\label{sec:size_distr}

\begin{figure}
\begin{center}
\includegraphics[width=8.5cm]{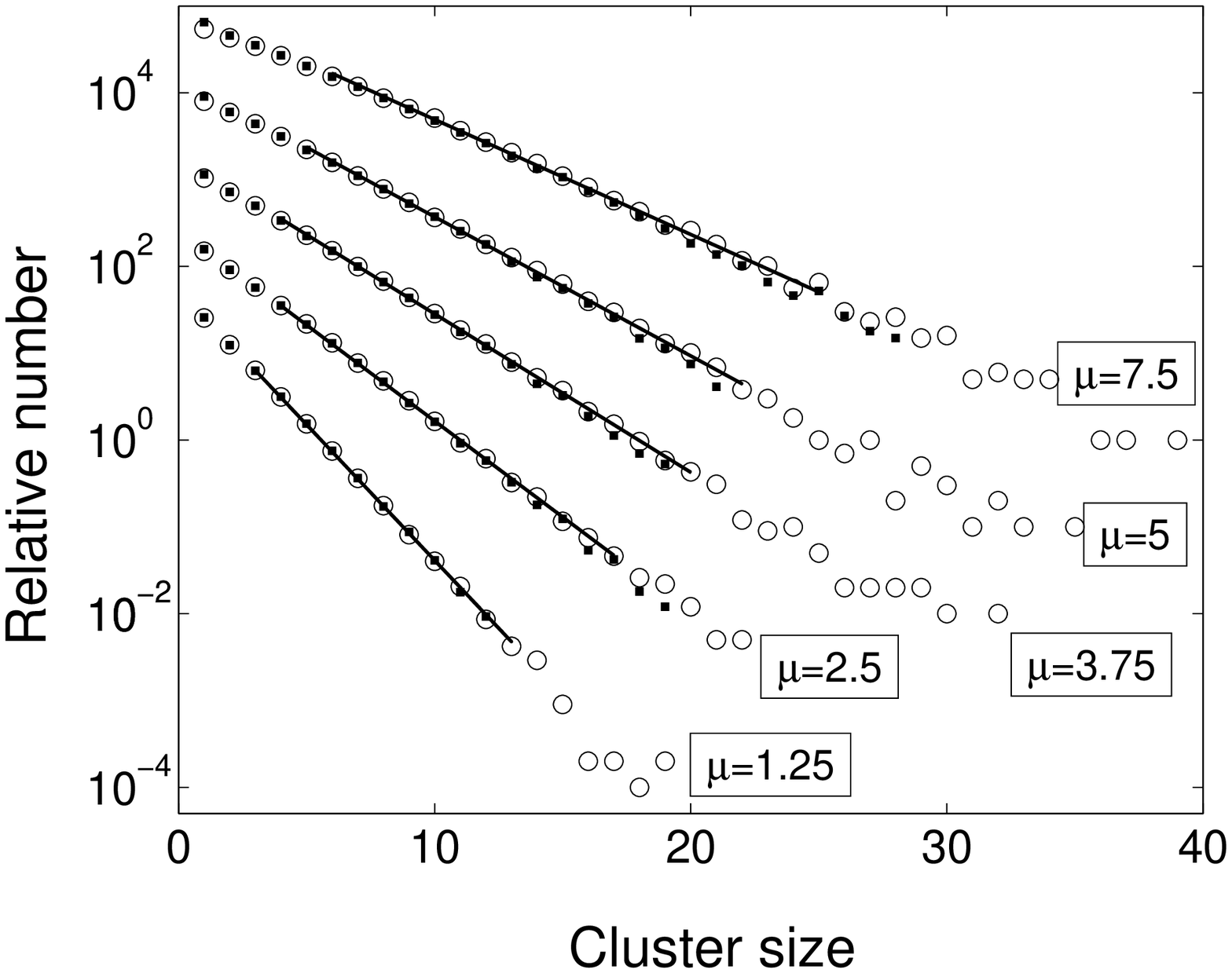}
\end{center}
\caption{\label{fig:cluster statistics mu} Number of clusters over a
range of sizes. The plot compares RG predictions (open circles) with
numerical integrations of Eqs.~(\ref{eq:model_general}) (solid squares).
The curves are spread out for clarity by multiplying
each successive data set as $\mu$ increases by $0.1$ (a linear shift
in the semilog plot).  The solid
lines denote the fits to the RG data over the domain described in
the text.}
\end{figure}

The main measurable physical quantity that arises in our analysis is the distribution of
cluster sizes. In Fig.~\ref{fig:cluster statistics mu} we show the
distribution of cluster sizes for several widths $\mu$ of the
coupling distribution from the RG and simulations.  The RG was
performed on one single realization of $10^6$ oscillators, while
the simulations were performed on $100$ realizations of $10^4$
oscillator chains.  Since the characteristic cluster size is much
smaller then $10^4$, we combine results from all $100$
realizations and analyze them as a single chain of size $10^6$. As
can be seen in Fig.~\ref{fig:cluster statistics mu}, good
agreement
exists between the RG and the simulations. 
The apparent discrepancies at the largest cluster sizes are due to
statistical fluctuations resulting from the small number of such
clusters present in the $10^6$ oscillator ensemble.

For a given value of the coupling width parameter $\mu$,
the probability of finding a cluster
is expected to fall off rapidly with increasing cluster size. The
linear scaling in the semi-log plot suggests that distribution of
cluster sizes has the form of an exponential: $P(n) \propto
\exp{\left(-n/\xi\right)}$.  In Fig.~\ref{fig:cluster statistics
mu} we exhibit the linear fits to the RG data.
The fits were made between twice the characteristic size $\xi$
and the cluster size at which there are just $20$ occurrences of
clusters of that given size, since for large clusters the scatter
becomes too large.  

\subsection{Dependence of $\xi$ on disorder strength}
\label{sec:xi_vs_mu}
The function $\xi(\mu)$ is a comprehensive quantity that
characterizes the physics of the chain in the unsynchronized
regime, because $\mu$ is the only control parameter in the chain,
and $\xi$ is the only quantity that characterizes the cluster
statistics since the distributions of cluster sizes are
exponential. We plot the function $\xi(\mu)$ in Fig.~\ref{fig:xi
versus mu}.
\begin{figure}
\begin{center}
\includegraphics[width=8.0cm]{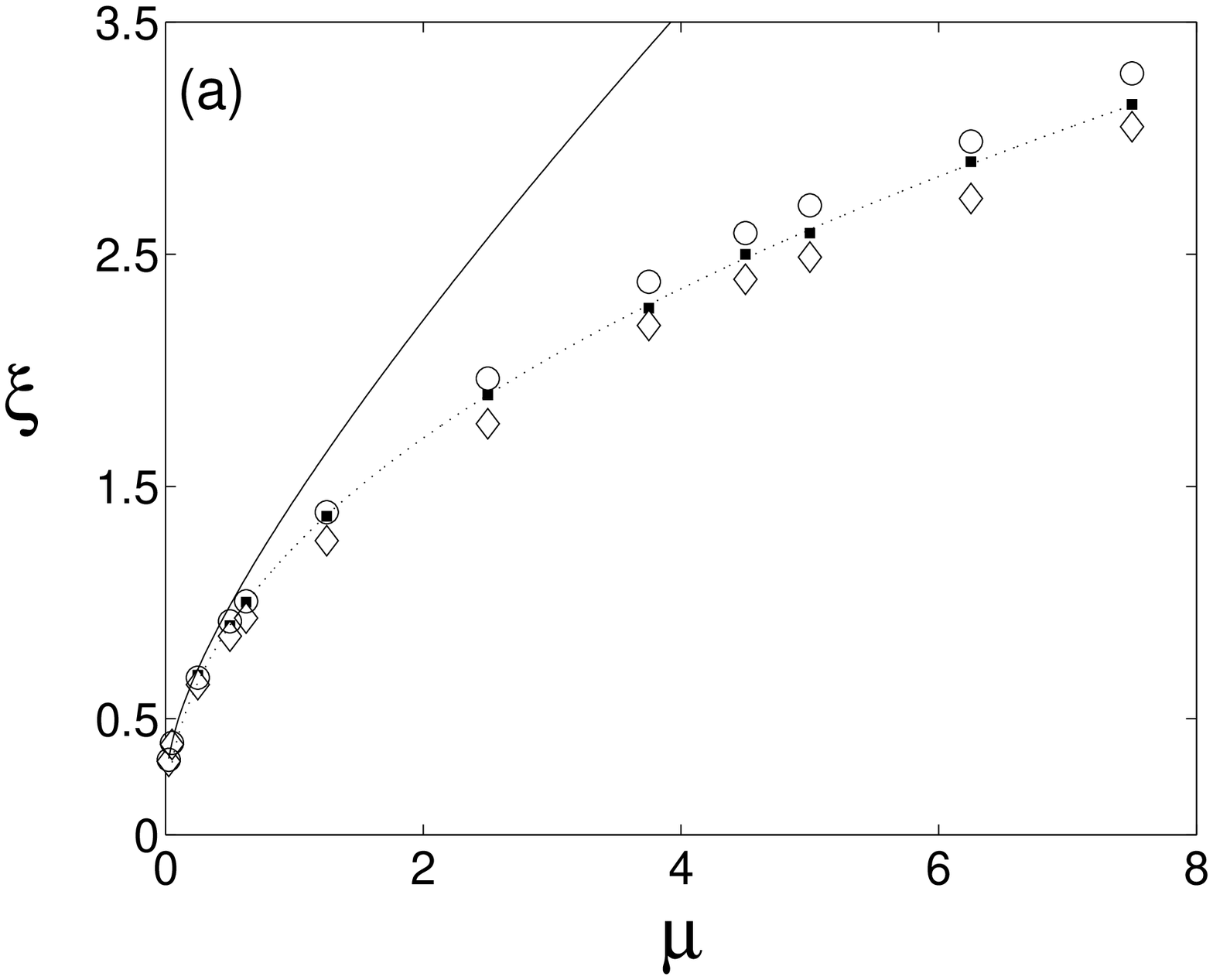}
\includegraphics[width=8.2cm]{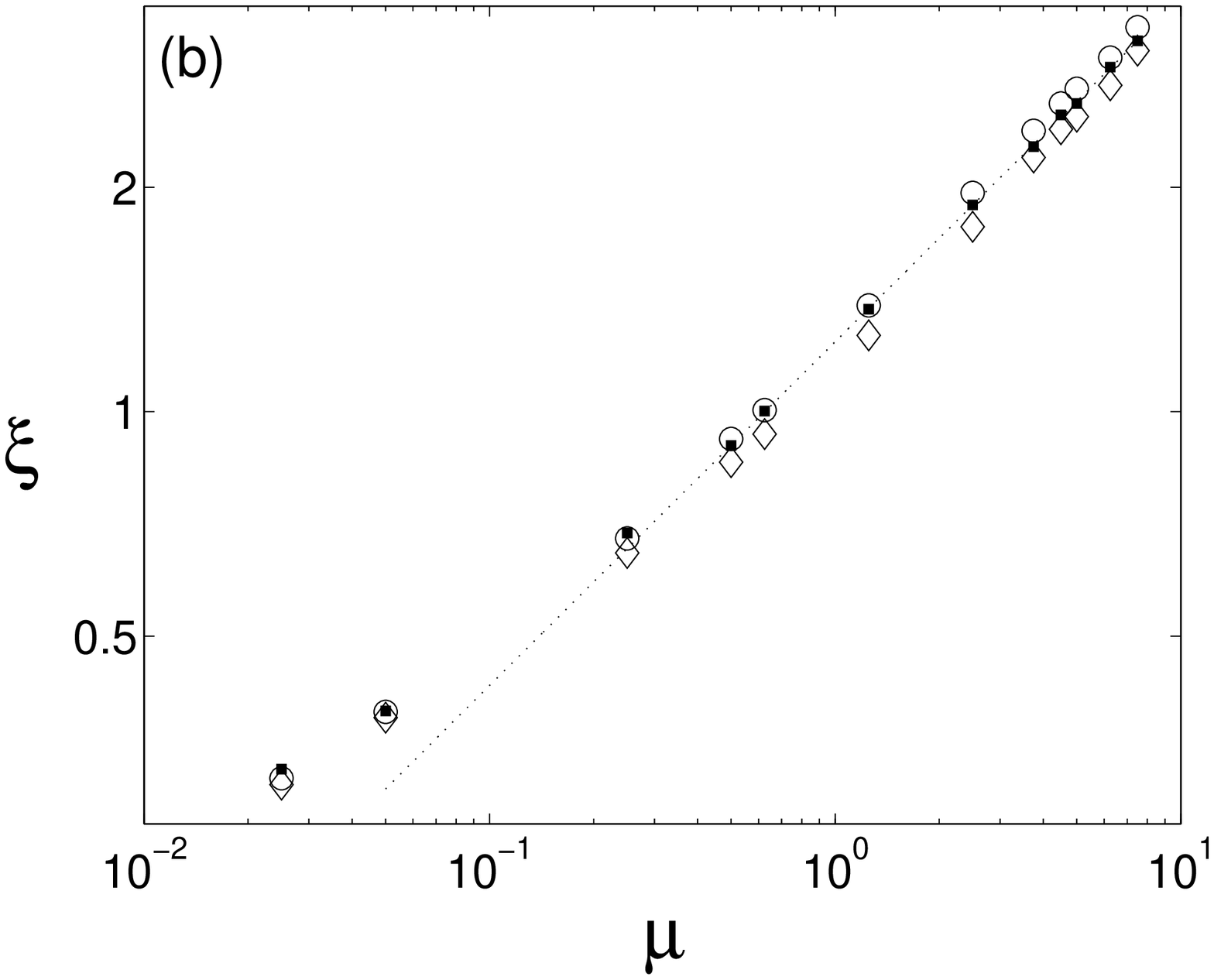}
\end{center}
\caption{\label{fig:xi versus mu} Characteristic cluster length
$\xi$ given by the fits to data in Fig.~\ref{fig:cluster
statistics mu}.  (a): Results of simulation (solid squares), the
numerical RG of Sec.~\ref{sec:Strong disorder RG} (open circles)
and an enhanced numerical RG with strain check (diamonds)
described in Sec.~\ref{sec:xi_vs_mu}. The solid line is a simple
analytic prediction $\xi(\mu)= -1/\log(1-\wp(\mu))$ where
$\wp(\mu)$ is defined in Eq.~(\ref{eq:wp}).  The dashed line is a
power law fit to the solid squares for $\mu > 0.625$ and has an
exponent of $0.46$; similar fits to the diamonds and circles give
exponents of $0.48$ and $0.48$ respectively. (b): Log-log plot of
the same data excluding the analytic prediction.}
\end{figure}
We find good agreement between the RG procedure (open circles) and
simulation (solid squares) until $\mu \approx 5$, after which the
two results begin to deviate from each other.  The small, but
growing discrepancy beyond $\mu = 5$ suggests there is
an additional mechanism in the physics of cluster formation
that becomes important at larger $\mu$
and is not taken into account by the decimation steps of the RG procedure.

A physical effect that may not be adequately included in our RG\
procedure is the accumulation of phase strain as a cluster is
built up through successive strong coupling decimation steps. We
can understand this effect based on the arguments of
Ref.~\cite{Strogatz and Mirollo 1}. If we neglect the interactions
of a synchronized cluster with the rest of the chain, its
frequency is the average frequency $\langle\omega\rangle$ of the
constituent oscillators. Furthermore, forming the sum of
Eqs.~(\ref{eq:model}) from one end of the cluster (here labeled 1)
to the $n$th
oscillator in the cluster gives%
\begin{equation}
\label{eq:str_chk}
X_{n}=\sum_{i=1}^{n}(\omega_{i}-\langle\omega\rangle)=K_{n}\sin(\theta
_{n+1}-\theta_{n}),
\end{equation}
since the internal interaction terms cancel. We refer to this condition for the
cluster's existence as a strain check. The condition reads:
$\left\vert X_{n}\right\vert <K_{n}$ for all $n$ in the cluster.
Note that $X_{n}$ is the sum of random numbers, so that as the
size of the cluster increases, the range of values of
$|X_{n}|$ typically grows, and increasing phase strains $|\theta_{n+1}%
-\theta_{n}|$ are needed to keep the various parts of the cluster
synchronized. If the strain becomes so large that some $|\theta_{n+1}%
-\theta_{n}|$ become $\pi/2$, the cluster will break. Also note
that if two clusters are combined through a strong coupling
decimation step to form a putative larger cluster, the mean frequency
$\langle\omega\rangle$ to be used in the sum of
Eq.~(\ref{eq:str_chk}) changes, and the condition $\left\vert
X_{n}\right\vert <K_{n}$ may be violated for a bond in the
interior of the two subclusters. This means
that the combined set of oscillators are not in fact synchronized,
and should be broken into smaller synchronized clusters in the RG
procedure. We proceed by making the assumption that the
oscillators form just two synchronized clusters, and identify the
break between the two clusters as where the condition $\left\vert
X_{n}\right\vert <K_{n}$ is violated in the presumed cluster.
Thus if $X_n$ exceeds $K_n$ at
some $n$, we do not allow the cluster to
form. Instead, we re-form the bare oscillators into
two different sub-clusters, such that the first sub-cluster is formed out
of the bare oscillators $1$ through $n$ and the second is formed
form the remaining bare oscillators.  The RG then
proceeds as normal.  The new sub-clusters may be subjected to
either of the two decimation steps later in the sweep, or in the next
sweep if more are required.

The $\xi(\mu)$ obtained with the version of the RG which included
the strain check are shown by diamonds in Fig.~\ref{fig:xi versus
mu}.  The change in the results takes place in the right direction
(decreasing $\xi$), but with a systematic over-estimation. 
All three sets of data, when plotted on a log-log scale follow
approximately straight lines for $\mu > 0.5$, so $\xi(\mu)$
appears to become a power law in this range.  
The straight-line fits to the data were made for $\mu > 0.625$.
The dashed curve in Fig.~\ref{fig:xi versus mu} is the fit to the
simulation results, suggesting the behavior $\xi \propto
\mu^{0.46}$ while the fits to the RG data with and without the
strain check give $\xi \propto \mu^{0.48}$ and $\mu^{0.48}$.

%

\subsection{Estimate of $\xi(\mu)$}
\label{sec:analytics} 

Given the simplicity of the renormalization group approach, one may
wonder whether we can capture the behavior of the oscillator chain
with a simple {\it one-loop} argument, based on the role of weak couplings in the limitation of
cluster sizes.

Indeed, the small clusters are likely to form between weak couplings,
and since the distribution of
separations between such weak couplings follows Poisson statistics, an
exponential distribution arises naturally.  Using this
hypothesis the probability of finding a cluster of size $n$ (i.e.,
a cluster delimited by two weak couplings $n$ units apart) is
$P(n,\wp) = C(1-\wp)^n = Ce^{-n/\xi}$ where $\xi =
-1/\ln{(1-\wp)}$ and $\wp$ is a probability that any randomly
chosen coupling is weak.  
We make an estimate of this $\wp$ based on bare distributions of $\omega$ and $K$.
Similarly to the RG, we define a coupling to be weak
if $2K < \left|\omega_{l} - \omega_{r}\right|$. For each such
bond, the probability that the required ratio is $>1$ is given by
\begin{equation}
\wp= 2\int_{0}^{\infty}\left(  \int_{-\infty}^{\infty}
g(\omega)g(\omega +\delta) \,d\omega\right)  \times\left(
\int_{0}^{\delta/2} G(K)\,dK\right)  \,d\delta
\end{equation}
The integral of $g(\omega)g(\omega+\delta)$ is the probability
that for each of the two oscillators connected by this bond,
$\omega_{l} - \omega_{r} = \delta$. For each such frequency
difference, all the bonds with $K < |\delta|/2$ are
considered weak.  
Finally, all possible values of $\delta$
are integrated in the outer integral. After performing the inner
integrals over $\omega$ and $K$ we obtain
\begin{equation}
\wp(\mu) = \frac{4}{\pi^{3}} \int_{0}^{\infty} \frac{2\pi}{4 +
\delta^{2}} \arctan{\left(  \frac{\delta}{2\mu}\right)} \,d\delta.
\label{eq:wp}
\end{equation}
This result 
is compared with both
the RG predictions and numerical solutions in Fig. \ref{fig:xi
versus mu}.  The expression gives a good description of the data for
$\mu \rightarrow 0$, but overestimates the characteristic cluster size for
larger values of $\mu$. The renormalization group method, which
does not treat all scales of frequencies and couplings at once,
clearly provides better predictions then an estimation based on
bare distributions.

\section{Conclusions}
\label{sec:Future+conclusion} In this paper we investigated
the collective behavior of a one-dimensional chain of coupled
non-linear oscillators with random frequencies and nearest-neighbor
couplings. For this purpose we developed a real space
renormalization group approach, which is expected to be
reliable in analyzing chains with large disorder. As we report above,
our RG approach did an excellent job in capturing both the dynamics of
individual oscillators, and the large scale behavior of the chain.

The disordered oscillator chain has the potential of establishing
macroscopic order, which is exhibited by a finite fraction of all
oscillators in the system moving in accord with the same
frequency. Indeed in systems with higher connectivities, such
global synchronization may occur, but in the limit of short ranged
interactions, the macroscopic order is stymied by fluctuations.
This dynamical behavior is reminiscent of critical phenomena in
equilibrium statistical mechanics, where for each system there
exists a lower critical dimension, and only dimensionality higher
than that allows macroscopic ordering. In our analysis, we
concentrated on the quantities that most directly capture the
collective aspects of the chain's behavior, and therefore also the
competition between the interactions which seek to establish a
synchronized phase, and the disorder-induced fluctuations which
destroy it. This physics is fully contained in the cluster-size
distribution, $P(n)\sim e^{-n/\xi(\mu)}$, which we find as a
function of the interaction-strength tuning parameter, $\mu$.

\begin{center}
\begin{figure}
\includegraphics[width=7.5cm]{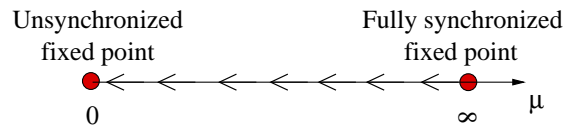}
\caption{Fixed points and flow diagram of the oscillator chain.
The short-range oscillator chain has only two fixed points: the
unstable fully synchronized fixed point at infinite interactions
(or zero disorder) $\mu\to\infty$, and the stable unsynchronized
fixed point. The cross over flow between the two points is
associated with a correlation length, captured by the average
cluster size, $\xi(\mu)$. \label{pd}}
\end{figure}
\end{center}

The behavior of the average cluster size, $\xi(\mu)$, acts as an
effective correlation length for the oscillator chain, where we
once more think of our system in terms of the theory of critical
phenomena. Since our oscillator chain is below its lower critical
dimension, it's phase diagram resembles that of the
one-dimensional Ising model, with $\mu$ serving the role of
inverse temperature $1/T$. The chain has an ordered phase, where
all oscillators are synchronized, which is reached in the limit
$\mu\to\infty$ (analogous to the Ferromagnetic phase at $T\to 0$).
This phase, however, is described by an {\it unstable} fixed
point. If an RG flow can be extended to the entire parameter
space, the full-synchronization fixed point flows under RG to the
stable fixed point found at the non-interacting point, $\mu=0$,
which describes the unsynchronized phase (analogous to the
paramagnetic fixed point at $T\to \infty$). The putative flow
diagram of the chain is shown in Fig. \ref{pd}. The flow from the
fully-synchronized point to the unsynchronized $\mu=0$ point is
indeed associated with a length scale, whose meaning is the
coarse-graining length at which the chain seems fully
unsynchronized, {\it i.e.}, consisting of single free oscillators.
This length is the size of synchronized clusters in the original
chain, $\xi(\mu)$.
Since $\xi(\mu)$ serves as an effective correlation length, it may
diverge in a universal fashion in the limit $\mu\to\infty$. While
our RG method can only access moderate values of $\mu$, we
cautiously argue that we can access the critical region, as can be
seen in Fig. \ref{fig:cluster statistics mu}. From our results we
obtain the critical exponent $\nu$ of the synchronization
transition: \be \xi(\mu)\sim \mu^{\nu} \hspace{1cm} \nu \simeq
0.48 \label{nu} \ee Note that this behavior is not analogous to
the divergence of the Ising chain correlation length $\xi \sim
\ln\frac{1}{T}$.

The method by which we obtained the above results constitutes the main
achievement of this work. We developed a real space renormalization
group scheme that successfully predicts the detailed behavior of a nonlinear
oscillator chain with short-range interactions.
We have implemented the RG scheme numerically on a chain of
$10^6$ oscillators and compared its results with essentially exact simulations of the model equations,
Eqs.~(\ref{eq:model}) (see Sec. \ref{sec:Results}).  The agreement is very good, and for some
characteristics, such as the cluster frequency distributions at
low frequencies, it improves with larger cluster sizes. This indicates
particularly that our local RG scheme successfully captures the non-trivial collective features of the
chain expressed in large synchronized clusters.

%

One practical implication of our analysis may be found when
considering coupled laser arrays, where only limited connectivity
can be achieved. Diode lasers, for example, are manufactured in
one-dimensional bars that allow for evanescent coupling through
overlapping electric fields. Due to the exponential fall-off in
the electric field envelope with distance, the coupling between
well-behaved diodes can only be short ranged. The tools developed
here could be used to study the extent of coherence that such
systems exhibit.  A potentially promising area of current research
is constructing grids of diodes coupled in 2-dimensions by
stacking 1-dimensional bars. Our RG technique could possibly also
be extended to such a system, and help clarify both its
properties, and whether it can establish macroscopic coherence.

Our work suggests several other directions for future research.
First, we may ask how can one improve the applicability of our RG
scheme to larger values of the interaction parameter $\mu$. This
may be pursued by focusing on the concept of intra-cluster strain,
which is roughly the variance of phase differences between
neighboring oscillators within a cluster. As interaction
increases, cluster sizes grow, and bigger strain is needed to
accommodate the spread of frequencies within a cluster (cf. Sec.
\ref{sec:xi_vs_mu}). In our current scheme we made the first
attempt to take this into account with the strain check based on
the argument of~\cite{Strogatz and Mirollo 1}, a condition
expressed by Eq.~(\ref{eq:str_chk}). This led to some improvement
- for example, $\xi(\mu)$ was modified in the right direction, but
this modification was over-estimated. The strain check may be
augmented in the future.

A second natural question is the degree of universality in the
oscillator chain. Our result for the correlation length critical
exponent $\nu$, Eq. (\ref{nu}), was obtained using Lorentzian
distributions. It is possible that the exponent depends on the nature
of the tails of the distributions used; Lorentzians, for instance, do
not have a variance. We intend to investigate the universality of
$\nu$ alongside investigating the applicability of our RG scheme to
distributions with narrower tails.

A more general question concerns the development of an RG scheme
and the analysis of synchronization in short-range networks at
higher dimensions, and especially at two dimensions, which is very
close to the lower critical dimension: Refs.~\cite{Anamalous,
Daido} suggest that in two dimensions there may be a transition to
a macroscopically synchronized phase, or that $d=2$ is  the lower
critical dimension giving rise to interesting anomalous behavior
(see also~\cite{Acerbon's_review}). The development of a
decimation procedure at dimensions higher than one is quite
challenging because of the higher connectivity of the system, and
the possibility that real-space local decimation steps change the
geometry of the system.



It is a pleasure to thank Heywood Tam for numerous discussions and
Tony Lee for carefully checking the RG procedure.  We are also
grateful to Boeing and the National Science Foundation under Grant
No.~DMR-0314069 for funding this work.


\end{document}